\documentclass[twocolumn,showkeys,prd,nofootinbib,preprintnumbers,floatfix]{revtex4-1}

\usepackage{amsfonts,amsmath,amssymb} 
\usepackage{graphicx,graphics,color}
\usepackage{gensymb}
\usepackage{longtable}
\usepackage{bbding}
\usepackage{hyperref}

  \def\prd{Phys. Rev. D}
 \def\prl{Phys. Rev. Lett.} 
  \def\physrep{Phys. Rep.}

\begin{document}

\title{Stability Conditions in the Generalized SU(2) Proca Theory}

\author{L.~Gabriel G\'omez}
\email{luis.gomez@correo.uis.edu.co}
\affiliation{Escuela de F\'isica, Universidad Industrial de Santander,\\ Ciudad Universitaria, Bucaramanga 680002, Colombia}

\author{Yeinzon Rodr\'{\i}guez}
\email{yeinzon.rodriguez@uan.edu.co}
\affiliation{Escuela de F\'isica, Universidad Industrial de Santander,\\ Ciudad Universitaria, Bucaramanga 680002, Colombia}
\affiliation{Centro de Investigaciones en Ciencias B\'asicas y Aplicadas, Universidad Antonio Nari\~no, \\
Cra 3 Este \# 47A-15, Bogot\'a D.C. 110231, Colombia}
\affiliation{Simons Associate at The Abdus Salam International Centre for Theoretical Physics, \\
Strada Costiera 11, I-34151, Trieste, Italy}


\preprint{PI/UAN-2019-651FT}

\begin{abstract}
Under the same spirit of the Galileon-Horndeski theories and their more modern extensions, the generalized SU(2) Proca theory was built by demanding that its action may be free of the Ostrogradski's instability. Nevertheless, the theory must also be free of other instability problems in order to ensure its viability. As a first approach to address this issue, we concentrate on a quite general variant of the theory and investigate the general conditions for the absence of ghost and gradient instabilities in the tensor sector 
without the need for resolving the dynamical background. 
The phenomenological interest of this approach as well as of the variant investigated lies on the possibility of building cosmological models driven solely by non-Abelian vector fields that may account for a successful description of both the early inflation and the late-time accelerated expansion of the universe.
\end{abstract}
\pacs{Valid PACS appear here}
\keywords{}
\maketitle


\section{Introduction}
General Relativity (GR) is the most successful 
theory of gravity to date.  On one hand, the theory has accounted for many fascinating phenomena in the local universe with extraordinary precision \cite{Akiyama:2019cqa,TheLIGOScientific:2017qsa,Monitor:2017mdv,GBM:2017lvd,Goldstein:2017mmi,Abuter:2018drb}; on the other hand, its degree of predictability makes it very appealing in light of future observations. 
The standard cosmological model, a remarkable achievement of both cosmologists and astronomers, has GR as its central pillar; however, it lacks of a convincing explanation for the current accelerated expansion of the universe driven by the, so called, dark energy \cite{Amendola:2015ksp}: that weird energy that keeps galaxies moving away from one another in an accelerated way at large scales and that comprises almost $70\%$ of the energy budget of the universe. Within the simple realization of the standard cosmological model, dark energy is assumed to be the cosmological constant $\Lambda$. There exists, nevertheless, an enormous discrepancy - some $120$ orders of magnitude \cite{Amendola:2015ksp} - between the value predicted by quantum field theory  and the one observed. 
Such a discrepancy has encouraged many theoreticians to propose alternative theories of gravity \cite{Clifton:2011jh,Tsujikawa:2010zza}, including quantum versions of it \cite{Rovelli:2004tv,Ashtekar:2004eh}; even so, the dark energy problem continues to be an unsolved problem. 

An interesting proposal, the so-called Galileon theory, was constructed with a scalar field as a gravitational degree of freedom in addition to the geometry. This scalar field enjoys a Galilean shift symmetry and derivative self interactions
\cite{2009PhRvD..79f4036N,2009PhRvD..79h4003D}. The generalization of the covariant Galileon was reported shortly after in the literature \cite{2009PhRvD..80f4015D,Deffayet:2011gz} as well as its vector counterpart, framed in what is called the generalized Proca theory \cite{Heisenberg:2014rta,Tasinato:2014eka,Allys:2015sht,Jimenez:2016isa,Allys:2016jaq}. The generalized Galileon theory is equivalent \cite{Kobayashi:2011nu} to the Horndeski's theory proposed decades ago \cite{Horndeski:1974wa}.  In contrast, the generalized Proca theory contains that formulated by Horndeski himself a coupled of years later \cite{Horndeski:1976gi} and that deals with an Abelian gauge field in curved spacetime.  Both kinds of theories lead to second-order field equations in a four dimensional spacetime, this being a necessary condition to avoid the Ostrogradski instability \cite{Ostrogradsky:1850fid} (see also Refs. \cite{Woodard:2006nt,Woodard:2015zca}).  Avoiding this instability has served as a guiding principle to formulate healthy extensions of GR. We refer the reader to the work in Ref. \cite{Kobayashi:2019hrl} for a comprehensive and recent review of scalar-tensor theories of gravity and to Ref. \cite{2019PhR...796....1H} for a general systematic approach in building field theories of gravity based on additional
scalar, vector, and tensor fields, as well as for an exploration of their cosmological and astrophysical applications. 

Further developments of this topic 
include the generalized Proca theory endowed with a global SU(2) symmetry \cite{2016PhRvD..94h4041A} (see also Ref. \cite{Jimenez:2016upj}). An interesting aspect of this theory is that the configuration of the non-Abelian vector fields is compatible with isotropic spacetimes due to the local isomorphism between the SO(3) and SU(2) groups of transformations. Such a configuration is described by a set of three space-like vector fields mutually orthogonal and of the same norm \cite{ArmendarizPicon:2004pm}. Thus, this cosmic triad turns out to be a natural arrangement inspired by successful ideas from particle physics.

Some interesting cosmological scenarios have been proposed based indeed on such a configuration for the vector fields \cite{Maleknejad:2011jw,2011PhRvD..84d3515M,Nieto:2016gnp,Adshead:2017hnc,2012PhRvL.108z1302A,Adshead:2016omu,Alvarez:2019ues}\footnote{See also the interesting proposal of Ref. \cite{Emami:2016ldl} that considers a triad of space-like vector fields charged under the global U(1) $\otimes$ U(1) $\otimes$ U(1) group of transformations.}. 
However, with the exception of  Refs. \cite{Rodriguez:2017ckc,Rodriguez:2017wkg} and some other works in progress \cite{infprep,beyondsu2}, nothing has been done in the context of the generalized SU(2) Proca theory unlike its scalar and Abelian vector analogues. 

On the other hand, stability aspects shape a fundamental issue for the construction of theoretically consistent models. This is then an inexorable examination one must perform for any gravity theory, even more if it involves extra gravitational degrees of freedom. For this reason, we concern about the stability criteria of the generalized SU(2) Proca theory. Thus, as a first approach to address this complex issue, we will focus on the tensor sector and investigate the general conditions for the avoidance of ghost and gradient instabilities
for a quite general variant of 
the generalized SU(2) Proca Lagrangian. 
Namely, we will concentrate on the practical conditions that make the theory cosmologically viable. 

The content of the paper has the following structure: Section \ref{sec:2} will be devoted to a general description of the generalized SU(2) Proca theory; we will also present here our arguments to establish a quite general variant that satisfies recent constraints \cite{ErrastiDiez:2019ttn,ErrastiDiez:2019trb} and that could be useful to model the primordial inflation \cite{infprep} as well as the late accelerated expansion of the universe \cite{Rodriguez:2017wkg}. 
In section \ref{sec:3}, we will define the tensor perturbations for the metric and vector fields and construct the second-order action for the tensor modes; then, we will assess the conditions to avoid ghost and gradient instabilities paying particular attention to analytical conditions for the coupling constants and without restricting the vector field dynamics; numerical computations will also be performed to visualize the general parameter space of the theory. Finally, we will discuss our results and present general remarks and perspectives of this work in section \ref{sec:6}.
\section{The generalized SU(2) proca theory}\label{sec:2}

A systematic approach to generalize the non-Abelian Proca theory in curved spacetime was elaborated in Ref. \cite{2016PhRvD..94h4041A} by demanding both its action 
to lead to equations of motion involving only second-order derivatives and the theory to propagate the right number of degrees of freedom. 

In the following, we will present the general Lagrangian of the theory under study containing up to two first-order derivatives of the vector field $A_{\mu}^a$ and up to six contracted space-time indices \cite{2016PhRvD..94h4041A,Rodriguez:2017wkg}. Here Latin letters running from 1 to 3 and Greek letters running from 0 to 3 are respectively used for the SU(2) group and space-time indices. All possible quartic $\mathcal{L}_{4}$ terms of the theory can be properly rewritten as linear combinations of the following Lagrangians:
\begin{eqnarray}
    \mathcal{L}_{4}^{1} &\equiv& \frac{1}{4}(A_{b}\cdot A^{b})[S_{\mu}^{\mu a}S_{\nu a}^{\nu}-S_{\nu}^{\mu a}S_{\mu a}^{\nu}+A_{a}\cdot A^{a}R] \nonumber \\
    &&+\frac{1}{2}(A_{a}\cdot A_{b})[S_{\mu}^{\mu a}S_{\nu}^{\nu b}-S_{\nu}^{\mu a}S_{\mu}^{\nu b}+2A^{a}\cdot A^{b}R] \,, \nonumber \\
     \mathcal{L}_{4}^{2} &\equiv& \frac{1}{4}(A_{b}\cdot A_{b})[S_{\mu}^{\mu a}S_{\nu }^{\nu b}-S_{\nu}^{\mu a}S_{\mu}^{\nu b}+A^{a}\cdot A^{b}R] \nonumber \\
     &&+\frac{1}{2}(A^{\mu a}A^{\nu b})[S_{\mu a}^{\rho}S_{\nu\rho b}-S_{\nu a}^{\rho}S_{\mu\rho b}-A_{a}^{\rho} A_{b}^{\sigma}R_{\mu\nu\rho\sigma} \nonumber \\
     &&-(\nabla^{\rho} A_{\mu a})(\nabla_{\rho} A_{\nu b})+ (\nabla^{\rho} A_{\nu a})(\nabla_{\rho} A_{\mu b})] \,, \nonumber \\
     \mathcal{L}_{4}^{3} &\equiv& \tilde{G}_{\mu\sigma}^{b} A_{a}^{\mu}A_{\nu b}S^{\nu\sigma a} \,,\label{sec:2:eqn1}
\end{eqnarray}
where $S_{\mu\nu}^{a}\equiv\nabla_{\mu}A_{\nu}^{a}+\nabla_{\nu}A_{\mu}^{a}$, $\tilde{G}_{\mu\nu}^{a} \equiv \frac{1}{2}\epsilon_{\mu\nu\rho\sigma}G^{\rho\sigma a}$ is the Hodge dual of $G_{\mu\nu}^{a}\equiv\nabla_{\mu}A_{\nu}^{a}-\nabla_{\nu}A_{\mu}^{a}$, $\nabla_\mu$ represents a space-time covariant derivative, $R$ is the Ricci scalar, and $R_{\mu \nu \rho \sigma}$ is the Riemman tensor. 
It was found quite recently, however, that theories involving multiple vector fields propagate unphysical degrees of freedom unless the condition
\begin{equation}
\frac{\partial^2 \mathcal{L}^{\rm Flat}_n}{\partial \dot{A}_0^a \partial A_0^b} - \frac{\partial^2 \mathcal{L}^{\rm Flat}_n}{\partial \dot{A}_0^b \partial A_0^a} = 0\,,
\end{equation}
is satisfied besides the usual vanishing of the determinant of the Hessian matrix \cite{ErrastiDiez:2019ttn,ErrastiDiez:2019trb}. In the previous expression, $\mathcal{L}^{\rm Flat}_n$ is the flat space-time version of $\mathcal{L}_n$. Such a condition has an important impact on the generalized SU(2) Proca theory since the pieces $\mathcal{L}_4^2$ and $\mathcal{L}_4^3$ are not allowed anymore. 

Extra terms corresponding exclusively to non-minimal couplings with gravity were also found in Ref. \cite{2016PhRvD..94h4041A} but we shall consider only the one involving four fields contracted with the double dual Riemann tensor $L^{\alpha\beta\gamma\delta} \equiv -\frac{1}{2}\epsilon^{\alpha\beta\mu\nu}\epsilon^{\gamma\delta\rho\sigma}R_{\mu\nu\rho\sigma}$:
%
\begin{equation}
      \mathcal{L}_{4}^{\rm Curv} \equiv L_{\mu\nu\rho\sigma}A^{\mu a}A^{\nu b}A_{a}^{\rho}A_{b}^{\sigma} \,.\label{sec:2:eqn2}
\end{equation}
$\mathcal{L}_4^1$ has a protagonist role in possible scenarios for dark energy \cite{Rodriguez:2017wkg} and primordial inflation \cite{infprep} so our reason for disregarding the other $\mathcal{L}^{\rm Curv}$ Lagrangian pieces found in Ref. \cite{2016PhRvD..94h4041A} is that $\mathcal{L}_{4}^{\rm Curv}$ is the only one whose structure is similar to that of $\mathcal{L}_4^1$:  both of them contain four vector fields.  

We will also consider $\mathcal{L}_{2}$ which corresponds to a minimal covariantization of the general flat space-time Lagrangian of a non-Abelian SU(2) gauge theory whose symmetry may be spontaneously broken:
\begin{equation}
    \mathcal{L}_{2} \equiv f(A_{\mu}^{a},F_{\mu\nu}^{a},\tilde{F}_{\mu\nu}^{a}) \,,\label{sec:2:eqn3}
\end{equation}
where $f$ is an arbitrary function of the vector field $A_\mu^a$, the gauge field strength tensor $F_{\mu \nu}^a$, and its Hodge dual $\tilde{F}_{\mu \nu}^a$;
in particular, we shall consider the standard Yang-Mills Lagrangian $\mathcal{L}_{\rm YM} \equiv -\frac{1}{4}F_{\mu\nu}^{a}F_{a}^{\mu\nu}$ as a  constituent part of the general model under study.  The Einstein-Hilbert term $\mathcal{L}_{\rm E-H} \equiv -\frac{R}{2}$, in physical units such that the reduced Planck mass $m_p=1$, will also be part of our study.  Thus, the action to explore is
\begin{equation}
   \mathcal{S}=\int d^{4}x\;\sqrt{-g} (\mathcal{L}_{\rm E-H}+   \mathcal{L}_{\rm YM}+ \alpha_1 \mathcal{L}_{4}^1+ \alpha_{\rm Curv} \mathcal{L}_{4}^{\rm Curv}) \,,\label{sec:2:eqn4}
\end{equation}
where $g$ is the determinant of the metric tensor and $\alpha_1, \alpha_{\rm Curv}$ are coupling constants. 

At this point, it is necessary to set the field configuration to describe completely the theory; for our purposes, we shall consider a homogeneous cosmic triad \cite{ArmendarizPicon:2004pm} which is fully compatible with a homogeneous and isotropic background. In short,
\begin{equation}
    A_{\mu}^{a}\equiv a(t)\phi(t) \delta_{\mu}^{a} \,,\label{sec:2:eqn5}
\end{equation}
where $\phi(t)$ is the physical vector field and $a(t)$ is the scale factor in the Friedmann-Lemaitre-Robertson-Walker (FLRW) metric whose line element in comoving Cartesian coordinates is $ds^{2}=-dt^{2}+a^{2}(t)\delta_{ij}dx^{i}dx^{j}$.

\section{Tensor Perturbations}\label{sec:3}
We consider the tensor perturbation in the metric tensor
\begin{equation}
\delta g_{ij}=a^{2}(t)h_{ij} \,,
\end{equation}
as well as for the perturbed SU(2) vector field
\begin{equation}
\delta A_{i}^{a}=a(t)t_{i}^{a} \,,
\end{equation}
where $h_{ij}$ and $t_{i}^{a}$ satisfy both the transverse and traceless conditions $\partial^{i}{h_{ij}}=h_{i}^{i}=0$ and $\delta_{a}^{i}\partial_{i}t_{j}^{a}=\delta_{a}^{i}t_{i}^{a}=0$, respectively. Hence, the tensor sector is expressed in terms of four dynamical modes associated to the polarization states $+$ and $\times$ for the metric and vector field as follows:
\begin{eqnarray}
   \delta g_{11} &=&-\delta g_{22}=a^{2}h_{+},\; \delta g_{12}=a^{2}h_{\times} \,, \nonumber \\
    \delta A_{\mu}^{1} &=& a(0,t_{+},t_{\times},0,0),\; \delta A_{\mu}^{2}=a(0,t_{\times},-t_{+},0,0) \,,
\end{eqnarray}
which can be oriented along the $z$-axis direction without loss of generality since modes of different momentum are not coupled to each other at this order in perturbations.

It is important to mention that, unlike theories not involving internal symmetries, e.g. Horndeski's theory and the generalized Proca theory, the inclusion of the global SU(2) symmetry in the vector field action renders the gravitational wave coupled to the vector field as a coupled oscillator. In particular when gravitational waves propagate through vector fields, their amplitudes are modulated by means of a redshift-dependent reduction. This effect is called gravitational wave-vector field oscillations \cite{2016PhRvD..94f3005C} and was recently studied in the gauge quintessence model \cite{Caldwell:2018feo} which is based on a SU(2) gauge field. Interestingly, this effect could be seen in future gravitational-wave observations. Hence, it will be  interesting to investigate in a future work such an effect in the framework of the generalized SU(2) Proca dark energy model \cite{Rodriguez:2017wkg} together with the restriction on the propagation speed of gravitational waves according to the LIGO/VIRGO observations \cite{TheLIGOScientific:2017qsa,Goldstein:2017mmi,GBM:2017lvd,Monitor:2017mdv}.

\subsection{Stability conditions for the tensor sector at the classical level}

Imposing 
the positiveness of the kinetic matrix and squared propagation speeds,
we shall be able to find some concrete relations and intervals of values for the coupling constants that constrain the theory. Such relations entail a suitable parameter space for the cosmological viability of the theory. In particular, we will focus as much as possible on those conditions that are valid for any dynamical behaviour of $\phi$, making this approach totally independent of the background dynamics.

\subsubsection{Ghost-free conditions}

We remind the reader that, in a dynamical system, the ghost-like instabilities corresponds to the presence of negative kinetic energy terms for the dynamical degrees of freedom of the linearized perturbations\footnote{See a very interesting and pedagogical review about the nature and consequences of ghosts in Ref. \cite{Sbisa:2014pzo}.}. Thus, building up a ghost-free theory demands a positive-definite kinetic matrix. 
In general, physical modes exhibit a scale dependence whereby stability must be guaranteed both in the short-wavelength and in the
long-wavelength regime. This is an issue that modified theories of gravity are commonly faced with because of the introduction of additional gravitational degrees of freedom. Nevertheless, due to the plain form of the perturbed pieces of the generalized SU(2) Proca Lagrangian once the cosmic triad configuration and the FLRW background have been imposed, the tensor modes show no scale dependence whereby this aspect can be overlooked.

We start by writing out the quadratic kinetic action after integrating by parts the terms $\ddot{h}_{+,\times}$, containing the products of first-order time derivatives for the dynamical modes  $\dot{\vec{x}}^{T} \equiv (h_{+}, t_{+}, h_{\times}, t_{\times})$, as
\begin{equation}
    S_{K}^{2}=\int d^{3}x\; dt\; a^{3} \dot{\vec{x}}^{T} K\; \dot{\vec{x}} \,, 
\end{equation}
where $K$ is a $4\times4$ symmetric matrix whose dimension is determined by the number of degrees of freedom. The non-vanishing components are
\begin{eqnarray}
    K_{11}&=&K_{13}=\frac{1}{4}+\left(\frac{61\alpha_1}{8}-2\alpha_{\rm Curv}\right)\phi^{4} \,, \\
    K_{22}&=&K_{44}=1+(5\alpha_1)\phi^{2} \,, \\
    K_{12}&=&(-5\alpha_1+4\alpha_{\rm Curv})\phi^3 \,, \\
    K_{34}&=&K_{12} \,.
    \end{eqnarray}
Notice that we have written the elements of the kinetic matrix in a general fashion involving all the coupling constants. Such elements follow the simple form: $N\times$coupling constants$\times\phi^n$ where $N$ is a rational number and $2<n<4$ depending on the type of perturbation, i.e. purely metric $n=4$, purely vector field $n=2$ or mixed one $n=3$.
To ease the analysis we define the ratio between the coupling constants as $d\equiv \alpha_{\rm Curv} / \alpha_1$. 
We keep nonetheless the same notation for the scalar field $\phi$ for brevity.

We proceed now to compute the eigenvalues of the kinetic matrix resulting in two degenerate solutions which are associated with the two polarization states, one for the metric and the other for the vector field:
%
\begin{equation}
  \lambda_{\pm}= \frac{1}{16} \{10 + \alpha_1 \phi^2 [40 + (61 - 16 d) \phi^2] \pm \sqrt{\Lambda} \} \,,
\end{equation}
where the discriminant $\Lambda$ is:
\begin{eqnarray}
    \Lambda &=& 36 + \alpha_1 \phi^2 \{ 480 + 4 (-183 + 400 \alpha_1 + 48 d) \phi^2  \nonumber \\ 
    &&+ 16 \alpha_1  [95  + 16 d (-35 + 16 d)] \phi^4 + \alpha_1 (61 - 16 d)^2 \phi^6 \} \,. \nonumber \\
    &&
\end{eqnarray}
Let us first start this analysis by stating that $\lambda_{+} \geq \lambda_-$. 
Hence, the main focus of this analysis lies on the requirement of the positiveness of $\lambda_{-}$ which leads to 
%
\begin{equation}
\alpha_1>0, \;\frac{1}{16} (15 - \sqrt{435}) \leq d \leq   \frac{1}{16} (15 + \sqrt{435}) \,. \label{gfe}
\end{equation}
%
Other conditions, not reported here, 
are possible but they are complicated enough and put quite severe constraints on the dynamics of the scalar field.
Nevertheless, for the sake of generality, we will take into account all the possible conditions in the next Subsection in order to plot the final parameter space once the respective conditions for the absence of gradient instabilities are imposed. 


%

\subsubsection{Gradient-instability-free conditions}

We proceed now by examining the conditions for the absence of Laplacian/gradient instabilities as complementary conditions to those for a ghost-free theory. To do so, we seek for products of first-order spatial derivatives for pure and mixed degrees of freedom. Then, after expanding each piece of the generalized SU(2) Proca  Lagrangian up to second order in tensor perturbations and integrating by parts the terms $\partial^{2}{h}_{+,\times}$ and $\partial^{2}{t}_{+,\times}$, the action containing those contributions encoded in the Laplacian $L$ matrix gets expressed in the form
\begin{equation}
    S_{L}^{2}=\int d^{3}x dt\; (-a\;\partial\vec{x}^{T} L\; \partial\vec{x}) \,.
\end{equation}
The components of the symmetric Laplacian matrix follow the same structure encountered when analyzing the kinetic matrix $K$ since they are sourced by the same type of perturbations:
\begin{eqnarray}
    L_{11}&=&L_{13}=\frac{1}{4}+\left(\frac{81}{8} \alpha_1 \right)\phi^{4} \,, \\
    L_{22}&=&L_{44}=1+(5 \alpha_1)\phi^{2} \,, \\
    L_{12}&=&-(5\alpha_1) \phi^{3} \,, \\
    L_{34}&=&L_{12} \,.
    \end{eqnarray}
As for the $K$ matrix structure, these components can be clearly viewed as a deviation from the standard contributions of Einstein-Hilbert and Yang-Mills tensor modes,  which both propagate at  the speed of light. The physical reason for such modifications are, clearly, the self-interactions of the vector field and its non-minimal coupling to gravity.

We do not identify any contribution from $\mathcal{L}_{4}^{\rm Curv}$ to this action up to second order in perturbations. In addition, it is worth observing the coupling between the vector and metric tensor modes through the mixed terms $L_{12}$ and $L_{21}$ as happens for the kinetic matrix.

The squared propagation speeds follow the dispersion relation ${\det}(c_{\rm T}^{2} K-L)=0$ 
in a non trivial way. Hence we require that $c_{\rm T\pm}^{2}>0$ at high enough frequencies to ensure oscillating solutions of the linearized perturbation equations of the tensor modes. Thus, the exact expressions for the squared sound speeds are 
%
\begin{widetext}
\begin{eqnarray}
 c_{\rm T_\pm}^{2} &=& \{ \alpha_1 \phi^2 [-10 + (-71 + 8 d) \phi^2 - 5 \alpha_1 (31 + 24 d) \phi^4] \pm 2 [1 + [\alpha_1^2 \phi^6 (1 + 5 \alpha_1 \phi^2) [64 d^2 + (5 + 4 d)^2 \phi^2 \nonumber \\
&& + \alpha_1 (125 + 8 d (-75 + 254 d)) \phi^4]]^{1/2}] \}/ \{ -2 + \alpha_1 \phi^2 [-10 + (-61 + 16 d) \phi^2 +  \alpha_1 [-105 + 16 d (-15 + 8 d)] \phi^4] \} \,. \nonumber \\
&& \label{lie}
\end{eqnarray}
\end{widetext}
Unexpectedly, we find that if $\mathcal{L}_4^{\rm Curv}$ does not contribute, i.e. $d = 0$, one tensor mode yields $c_{\rm T}^{2}=1$, without imposing any condition. This appealing realization arises due to some conspiring cancellation of the tensor modes: the effects of modified gravity are somehow counteracted by those of the vector field perturbations. For the second mode we note that there is a minimum $c_{\rm T +}^{2}=1$ around $\phi=0$. However, for very small enough values of $\alpha_1$, this minimum value is also applicable to a much more ample spectrum of values for $\phi$.

Before proceeding, let us first make a short comment about another crucial issue at fundamental level. Aside from imposing the conditions for the absence of ghosts and Laplacian instabilities, one should be worried about superluminal propagation of the metric and field fluctuations which could, in principle, violate causality. Nevertheless, it has been shown that superluminal propagation not necessarily leads to the appearance of closed (time-like) causal curves \cite{Babichev:2007dw}. In other words, in spite of the presence
of superluminal signals, causality may be preserved due to non-trivial backgrounds and fields. We do not enter however in this discussion and prefer to tackle this subject separately in a future publication such as has been done, for example, for the bi- and multi-Galileon models \cite{2011PhRvD..83h5015G,deFromont:2013iwa}.

Turning now to the central discussion, Eq. (\ref{lie}) tells us that the dynamics-independent condition for the absence of gradient instabilities, i.e. for $c^{2}_{s_\pm}>0$, is weaker than that required for the absence of ghost instabilities and reported in Eq. (\ref{gfe}).
We note that there exist some other possibilities for $c^{2}_{s_\pm}>0$ which impose quite serious constraints on the $\phi$ behaviour. 
We anticipate the serious concern of whether those other possibilities are consistent with the dynamical evolution of the background. Tough we have assumed in this work a different approach to constrain this theory, i.e. without solving the dynamic evolution, we expect a successful scenario for either very small values of $\alpha_1$, which admit a larger parameter space in the plane $\phi,d$, 
or $\alpha_1\lesssim 1$ together with the constraint in Eq. (\ref{gfe}). These statements are better illustrated through numerical computations in the next section.

It is interesting to see that clear constraints on the coupling constants of the pieces of the generalized SU(2) Proca Lagrangian have been derived from a mere analysis of the  tensor sector, without specifying the background evolution. Thus far, we have gained valuable information on the coupling constants at relatively low cost due to the simple structure of the tensor modes at second order in perturbations at the level of the action.


\subsection{Numerical results}

\begin{figure}[!htb]
\centering
\includegraphics[width=0.7\hsize,clip]{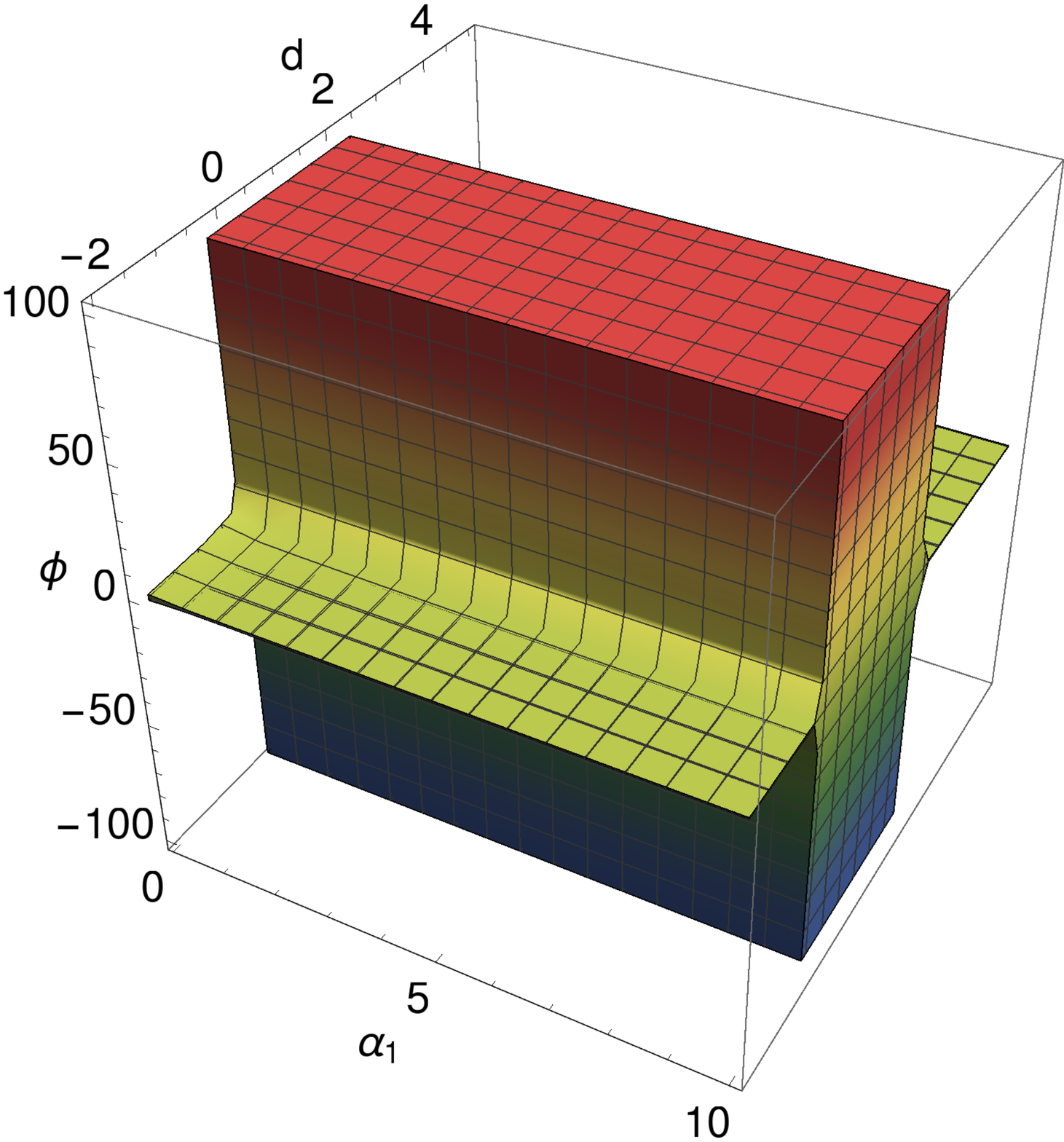}
\caption{Overall parameter space where the ghost-free/Laplacian-instability-free conditions are satisfied. 
}\label{fig:models23}
\end{figure}

After applying the criteria for the avoidance of ghost and gradient instabilities, we plot in Fig.~\ref{fig:models23} the resulting conditions 
that represent the suitable parameter space. We can readily see that the analytical results found before are contained in this plot and correspond to the general representation of the parameter space as we advertised. We also note that the conditions for the absence of ghost instabilities are more severe and hence dominate the overall parameter space.

\begin{figure}[!htb]
\centering
\includegraphics[width=0.7\hsize,clip]{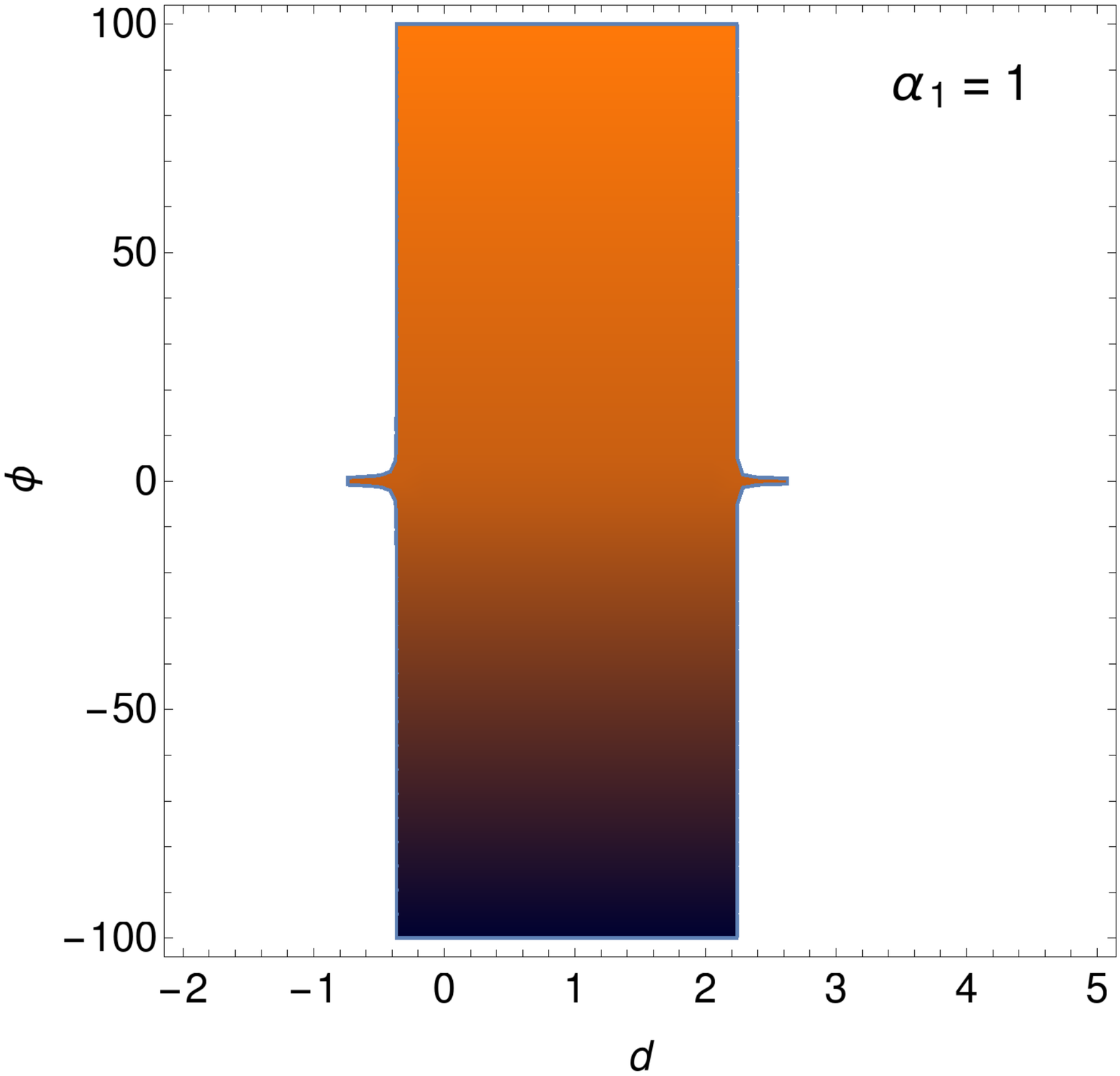}
\includegraphics[width=0.7\hsize,clip]{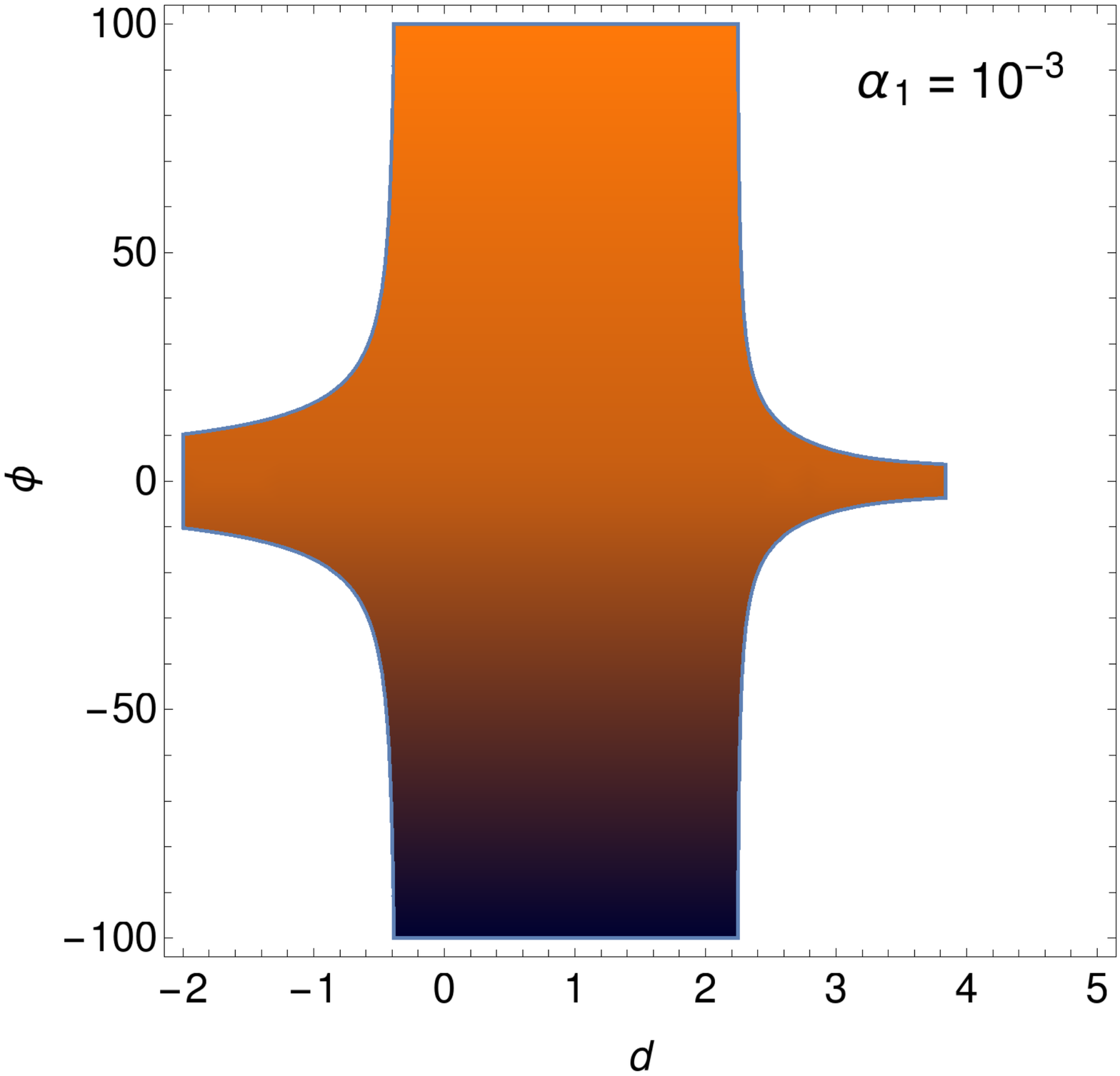}
\caption{Cross section of Fig. \ref{fig:models23} for $\alpha_1 = 1$ and $\alpha_1 = 10^{-3}$ respectively.  The available parameter space is larger in the $\phi,d$ plane for smaller values of $\alpha_1$.}\label{morefigs1}
\end{figure}

\begin{figure}[!htb]
\centering
\includegraphics[width=0.7\hsize,clip]{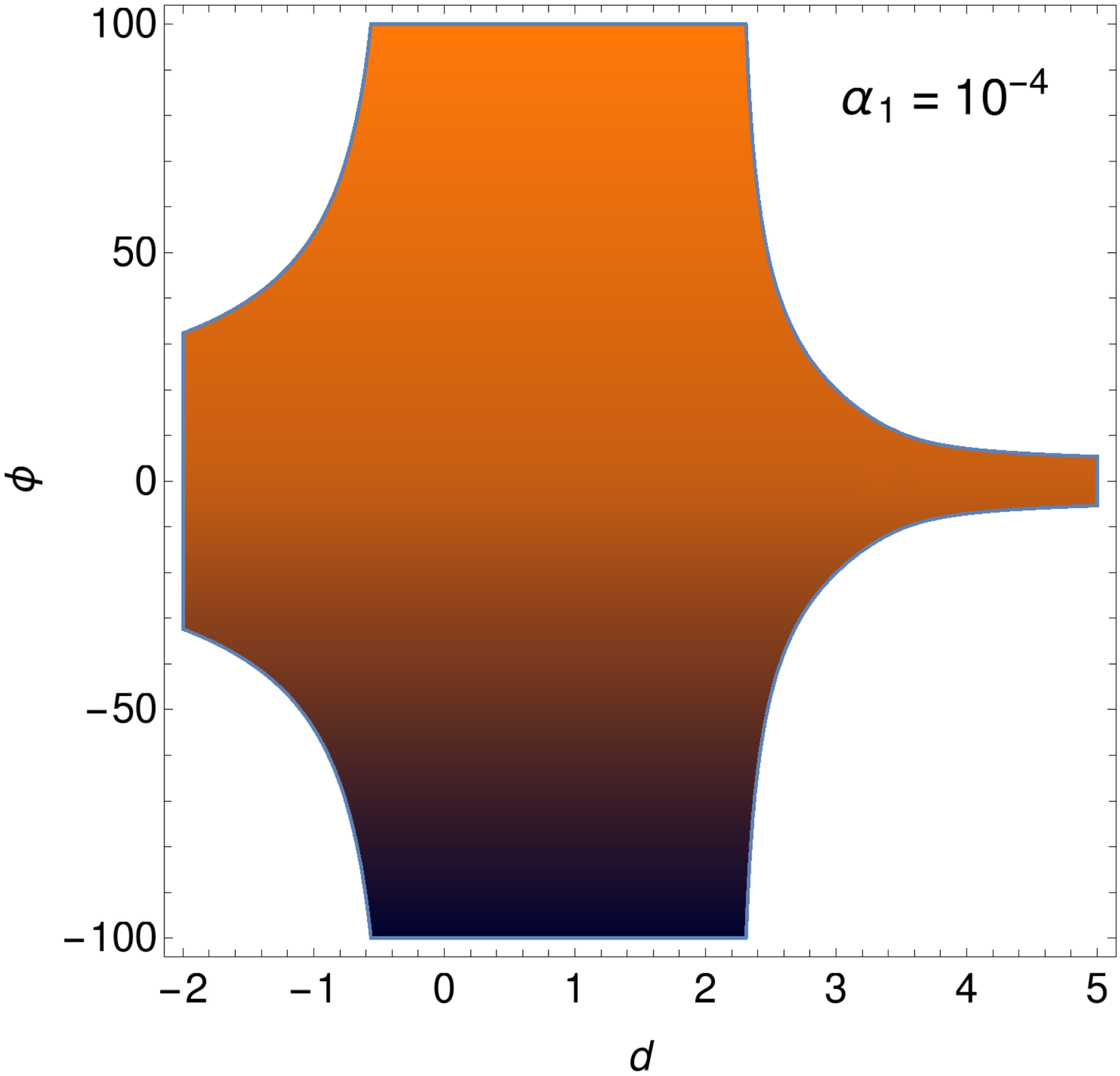}
\includegraphics[width=0.7\hsize,clip]{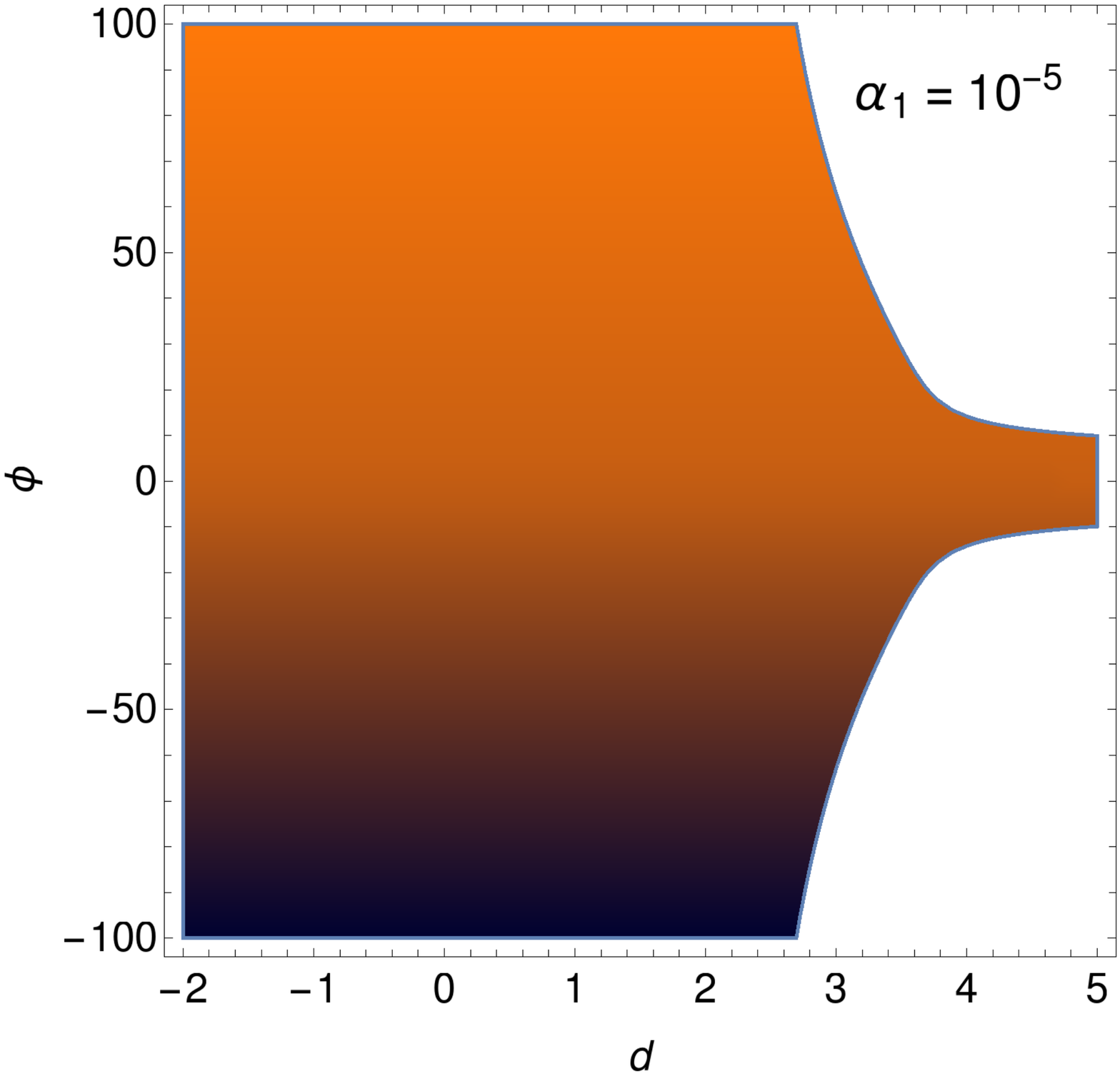}
\caption{Same as Fig. \ref{morefigs1} but with $\alpha_1 = 10^{-4}$ and $\alpha_1 = 10^{-5}$ respectively. }\label{morefigs2}
\end{figure}

As a general feature of this result, we note that those other solutions mentioned in the previous Subsection are encoded in the horizontal wedge centered at $\phi=0$ and their explicit expressions depend on cumbersome combinations of $\alpha_1,d$ and $\phi$.
Indeed, for very small values of $\alpha_1$, the parameter space is larger in the $\phi,d$ plane (see Figs. \ref{morefigs1} and \ref{morefigs2}). That is to say that keeping the dynamics of the vector field rather free, the theory turns out to be phenomenologically appealing.

\subsubsection{Asymptotic limit for $\phi$}

We speculate that the energy density associated to $\phi$ can strongly dominate as the universe evolves so as
to produce a significant accelerated expansion. Thereby we are allowed to assume that the field  may play the distinguished role of the dark energy \cite{Rodriguez:2017wkg} or of the inflaton field \cite{infprep}. Let us consider as an useful approach the asymptotic value of the vector field $\phi\to\pm\infty$ once the universe has evolved. As a consequence, we find the reduced form  $\lambda_{\pm}=\eta\pm|\eta|$ for the eigenvalues and the reduced expressions for the squared propagation speeds
as follows
%
\begin{eqnarray}
\lambda_{\pm} &=& \alpha_1 (61-16d)\pm |\alpha_1 (61-16d)| \,, \\
c_{\rm T\pm}^{2} &=& \frac{5 \alpha_1^2 (31 + 24 d) + 2 \sqrt{\Xi}}{\alpha_1^2 (105 + 240 d - 128 d^2)} \,,
\end{eqnarray}
 %
with 
\begin{equation}
\Xi \equiv 5\alpha_1^4 (125 - 600 d + 2032 d^2) \,,
\end{equation}
%
We can observe the impossibility of having simultaneously $\lambda_{\pm}>0$  whereby it is necessary to relax one (necessary) condition and set $\lambda_{-}=0$. This result, of course, does not guarantee a positive kinetic energy for the metric tensor modes.  Therefore, a more robust condition must be added.
Accordingly, the parameter space is 
dominated as before by the conditions that guarantee the absence of ghost instabilities, namely 
$-0.37 \lesssim d \lesssim 2.24$.

One can check as a simple examination that when $\phi\to 0$, the Einstein Yang-Mills theory is recovered. This setup entails 
$\lambda_{-}=1/4$ and $\lambda_{+}=1$ for the gravitational field and gauge field respectively, with the two tensor modes propagating at the speed of light $c_{\rm T \pm}^{2}=1$.

\section{Discussion and conclusions}\label{sec:6}

A fundamental issue that any theory of gravity has to face is the possible presence of instability problems known as ghost and gradient instabilities.
This issue, together with the gravity waves speed, is the most important criteria to constrain theories like Horndeski's or the generalized Proca at the fundamental level. In this paper, we have approached this delicate aspect aiming at inquiring about the cosmological viability of the generalized SU(2) Proca theory \cite{2016PhRvD..94h4041A}. As a first step, we have concentrated our attention on the tensor sector which, unlike its scalar and vector analogues, can be treated independently of the background dynamics under certain feasible conditions. To do so, we have implemented the following procedure:
first, we have proposed to study the action in Eq. (\ref{sec:2:eqn4}) 
assuming that it may play an important role in the evolution of the universe with no specification of the background evolution; second, we have found the general conditions for the absence of ghost and gradient instabilities in the tensor sector to render the theory phenomenologically viable; from here we have derived an analytical solution, Eq. (\ref{gfe}), that applies in most situations and that 
define the suitable parameter space in a simpler manner; 
and third, and for the sake of generality, we have numerically obtained the general conditions 
  and verified the analytical results previously found.
The asymptotic value of the vector field $\phi\to \pm \infty$ was always considered 
to examine in a clearer way the stability conditions.  Our results show that the absence of instabilities for the tensor perturbations already constrains in an important way the parameter space of the theory.  We will leave for a future work the analysis of the possible instabilities in the scalar and vector perturbations.

It is worth pointing out that, according to the constraints in Eq. (\ref{gfe}), the dark energy model of Ref. \cite{Rodriguez:2017wkg} is free of ghost and Laplacian instabilities.  This is because this model is identical to the one studied in this paper except for the $\mathcal{L}_4^{\rm Curv}$ contribution, which makes $d = 0$, requiring in turn $\alpha_1 > 0$ for the emergence of the late accelerated expansion mechanism.

Thus, we have provided an instructive guidance that can be implemented in cosmological scenarios based on SU(2) vector fields. It remains, however, to check whether our results are consistent with the LIGO/VIRGO observations \cite{TheLIGOScientific:2017qsa,Goldstein:2017mmi,GBM:2017lvd,Monitor:2017mdv} of gravitational waves for dark energy models. It will be also interesting to investigate the  effect of gravitational wave-vector field oscillations \cite{2016PhRvD..94f3005C,Caldwell:2018feo} in the framework of the generalized SU(2) Proca theory. This appealing feature opens up the possibility of new searches in light of future experiments on gravitational waves aiming at detecting new phenomena. 

Superluminal propagation of the fluctuations in the generalized SU(2) Proca theory is permitted as in many modified theories of gravity with non trivial backgrounds and extra degrees of freedom (see e.g. Ref. \cite{Babichev:2007dw}). Hence, causality is another unavoidable issue we will have to address in the future to ensure the cosmological viability of this theory at the fundamental level.

We can finally conclude that these results represent the first progress toward constraining this theory from observations and theoretical arguments and constitute an important insight for building cosmological models based on non-Abelian vector fields. 

\section{Acknowledgements}
We would like to offer our gratitude to Alex\'ander Ga\-lle\-go Cadavid for careful reading of an earlier version of this paper and to Juan Camilo Garnica for his continuous help, discussions, and verification of the results presented here.  Y.R. acknowledges Diana L\'opez-Nacir for an interesting discussion which helped clarify some aspects regarding instabilities. L.G.G. is supported by the  VIE - UIS Postdoctoral Fellowship Programme N$^\circ$ $	
2018000101$.  This work was supported by the following grants: Colciencias-DAAD - 110278258747 RC-774-2017, VCTI - UAN - 2017239, DIEF de Ciencias - UIS - 2460, and Centro de Investigaciones - USTA - 1952392.

\bibliography{apssamp}

\begin{thebibliography}{51}%
\makeatletter
\providecommand \@ifxundefined [1]{%
 \@ifx{#1\undefined}
}%
\providecommand \@ifnum [1]{%
 \ifnum #1\expandafter \@firstoftwo
 \else \expandafter \@secondoftwo
 \fi
}%
\providecommand \@ifx [1]{%
 \ifx #1\expandafter \@firstoftwo
 \else \expandafter \@secondoftwo
 \fi
}%
\providecommand \natexlab [1]{#1}%
\providecommand \enquote  [1]{``#1''}%
\providecommand \bibnamefont  [1]{#1}%
\providecommand \bibfnamefont [1]{#1}%
\providecommand \citenamefont [1]{#1}%
\providecommand \href@noop [0]{\@secondoftwo}%
\providecommand \href [0]{\begingroup \@sanitize@url \@href}%
\providecommand \@href[1]{\@@startlink{#1}\@@href}%
\providecommand \@@href[1]{\endgroup#1\@@endlink}%
\providecommand \@sanitize@url [0]{\catcode `\\12\catcode `\$12\catcode
  `\&12\catcode `\#12\catcode `\^12\catcode `\_12\catcode `\%12\relax}%
\providecommand \@@startlink[1]{}%
\providecommand \@@endlink[0]{}%
\providecommand \url  [0]{\begingroup\@sanitize@url \@url }%
\providecommand \@url [1]{\endgroup\@href {#1}{\urlprefix }}%
\providecommand \urlprefix  [0]{URL }%
\providecommand \Eprint [0]{\href }%
\providecommand \doibase [0]{http://dx.doi.org/}%
\providecommand \selectlanguage [0]{\@gobble}%
\providecommand \bibinfo  [0]{\@secondoftwo}%
\providecommand \bibfield  [0]{\@secondoftwo}%
\providecommand \translation [1]{[#1]}%
\providecommand \BibitemOpen [0]{}%
\providecommand \bibitemStop [0]{}%
\providecommand \bibitemNoStop [0]{.\EOS\space}%
\providecommand \EOS [0]{\spacefactor3000\relax}%
\providecommand \BibitemShut  [1]{\csname bibitem#1\endcsname}%
\let\auto@bib@innerbib\@empty
\bibitem [{\citenamefont {Akiyama}\ \emph {et~al.}(2019)\citenamefont {Akiyama}
  \emph {et~al.}}]{Akiyama:2019cqa}%
  \BibitemOpen
  \bibfield  {author} {\bibinfo {author} {\bibfnamefont {K.}~\bibnamefont
  {Akiyama}} \emph {et~al.} (\bibinfo {collaboration} {Event Horizon
  Telescope}),\ }\href {\doibase 10.3847/2041-8213/ab0ec7} {\bibfield
  {journal} {\bibinfo  {journal} {Astrophys. J.}\ }\textbf {\bibinfo {volume}
  {875}},\ \bibinfo {pages} {L1} (\bibinfo {year} {2019})},\ \Eprint
  {http://arxiv.org/abs/1906.11238} {arXiv:1906.11238 [astro-ph.GA]}
  \BibitemShut {NoStop}%
\bibitem [{\citenamefont {Abbott}\ \emph
  {et~al.}(2017{\natexlab{a}})\citenamefont {Abbott} \emph
  {et~al.}}]{TheLIGOScientific:2017qsa}%
  \BibitemOpen
  \bibfield  {author} {\bibinfo {author} {\bibfnamefont {B.~P.}\ \bibnamefont
  {Abbott}} \emph {et~al.} (\bibinfo {collaboration} {LIGO Scientific,
  Virgo}),\ }\href {\doibase 10.1103/PhysRevLett.119.161101} {\bibfield
  {journal} {\bibinfo  {journal} {Phys. Rev. Lett.}\ }\textbf {\bibinfo
  {volume} {119}},\ \bibinfo {pages} {161101} (\bibinfo {year}
  {2017}{\natexlab{a}})},\ \Eprint {http://arxiv.org/abs/1710.05832}
  {arXiv:1710.05832 [gr-qc]} \BibitemShut {NoStop}%
\bibitem [{\citenamefont {Abbott}\ \emph
  {et~al.}(2017{\natexlab{b}})\citenamefont {Abbott} \emph
  {et~al.}}]{Monitor:2017mdv}%
  \BibitemOpen
  \bibfield  {author} {\bibinfo {author} {\bibfnamefont {B.~P.}\ \bibnamefont
  {Abbott}} \emph {et~al.} (\bibinfo {collaboration} {LIGO Scientific, Virgo,
  Fermi-GBM, INTEGRAL}),\ }\href {\doibase 10.3847/2041-8213/aa920c} {\bibfield
   {journal} {\bibinfo  {journal} {Astrophys. J.}\ }\textbf {\bibinfo {volume}
  {848}},\ \bibinfo {pages} {L13} (\bibinfo {year} {2017}{\natexlab{b}})},\
  \Eprint {http://arxiv.org/abs/1710.05834} {arXiv:1710.05834 [astro-ph.HE]}
  \BibitemShut {NoStop}%
\bibitem [{\citenamefont {Abbott}\ \emph
  {et~al.}(2017{\natexlab{c}})\citenamefont {Abbott} \emph
  {et~al.}}]{GBM:2017lvd}%
  \BibitemOpen
  \bibfield  {author} {\bibinfo {author} {\bibfnamefont {B.~P.}\ \bibnamefont
  {Abbott}} \emph {et~al.} (\bibinfo {collaboration} {LIGO Scientific, Virgo,
  Fermi GBM, INTEGRAL, IceCube, AstroSat Cadmium Zinc Telluride Imager Team,
  IPN, Insight-Hxmt, ANTARES, Swift, AGILE Team, 1M2H Team, Dark Energy Camera
  GW-EM, DES, DLT40, GRAWITA, Fermi-LAT, ATCA, ASKAP, Las Cumbres Observatory
  Group, OzGrav, DWF (Deeper Wider Faster Program), AST3, CAASTRO, VINROUGE,
  MASTER, J-GEM, GROWTH, JAGWAR, CaltechNRAO, TTU-NRAO, NuSTAR, Pan-STARRS,
  MAXI Team, TZAC Consortium, KU, Nordic Optical Telescope, ePESSTO, GROND,
  Texas Tech University, SALT Group, TOROS, BOOTES, MWA, CALET, IKI-GW
  Follow-up, H.E.S.S., LOFAR, LWA, HAWC, Pierre Auger, ALMA, Euro VLBI Team, Pi
  of Sky, Chandra Team at McGill University, DFN, ATLAS Telescopes, High Time
  Resolution Universe Survey, RIMAS, RATIR, SKA South Africa/MeerKAT}),\ }\href
  {\doibase 10.3847/2041-8213/aa91c9} {\bibfield  {journal} {\bibinfo
  {journal} {Astrophys. J.}\ }\textbf {\bibinfo {volume} {848}},\ \bibinfo
  {pages} {L12} (\bibinfo {year} {2017}{\natexlab{c}})},\ \Eprint
  {http://arxiv.org/abs/1710.05833} {arXiv:1710.05833 [astro-ph.HE]}
  \BibitemShut {NoStop}%
\bibitem [{\citenamefont {Goldstein}\ \emph {et~al.}(2017)\citenamefont
  {Goldstein} \emph {et~al.}}]{Goldstein:2017mmi}%
  \BibitemOpen
  \bibfield  {author} {\bibinfo {author} {\bibfnamefont {A.}~\bibnamefont
  {Goldstein}} \emph {et~al.},\ }\href {\doibase 10.3847/2041-8213/aa8f41}
  {\bibfield  {journal} {\bibinfo  {journal} {Astrophys. J.}\ }\textbf
  {\bibinfo {volume} {848}},\ \bibinfo {pages} {L14} (\bibinfo {year}
  {2017})},\ \Eprint {http://arxiv.org/abs/1710.05446} {arXiv:1710.05446
  [astro-ph.HE]} \BibitemShut {NoStop}%
\bibitem [{\citenamefont {Abuter}\ \emph {et~al.}(2018)\citenamefont {Abuter}
  \emph {et~al.}}]{Abuter:2018drb}%
  \BibitemOpen
  \bibfield  {author} {\bibinfo {author} {\bibfnamefont {R.}~\bibnamefont
  {Abuter}} \emph {et~al.} (\bibinfo {collaboration} {GRAVITY}),\ }\href
  {\doibase 10.1051/0004-6361/201833718} {\bibfield  {journal} {\bibinfo
  {journal} {Astron. Astrophys.}\ }\textbf {\bibinfo {volume} {615}},\ \bibinfo
  {pages} {L15} (\bibinfo {year} {2018})},\ \Eprint
  {http://arxiv.org/abs/1807.09409} {arXiv:1807.09409 [astro-ph.GA]}
  \BibitemShut {NoStop}%
\bibitem [{\citenamefont {Amendola}\ and\ \citenamefont
  {Tsujikawa}(2015)}]{Amendola:2015ksp}%
  \BibitemOpen
  \bibfield  {author} {\bibinfo {author} {\bibfnamefont {L.}~\bibnamefont
  {Amendola}}\ and\ \bibinfo {author} {\bibfnamefont {S.}~\bibnamefont
  {Tsujikawa}},\ }\href
  {http://www.cambridge.org/academic/subjects/physics/cosmology-relativity-and-gravitation/dark-energy-theory-and-observations?format=PB&isbn=9781107453982}
  {\emph {\bibinfo {title} {{Dark Energy}}}}\ (\bibinfo  {publisher} {Cambridge
  University Press},\ \bibinfo {year} {2015})\BibitemShut {NoStop}%
\bibitem [{\citenamefont {Clifton}\ \emph {et~al.}(2012)\citenamefont
  {Clifton}, \citenamefont {Ferreira}, \citenamefont {Padilla},\ and\
  \citenamefont {Skordis}}]{Clifton:2011jh}%
  \BibitemOpen
  \bibfield  {author} {\bibinfo {author} {\bibfnamefont {T.}~\bibnamefont
  {Clifton}}, \bibinfo {author} {\bibfnamefont {P.~G.}\ \bibnamefont
  {Ferreira}}, \bibinfo {author} {\bibfnamefont {A.}~\bibnamefont {Padilla}}, \
  and\ \bibinfo {author} {\bibfnamefont {C.}~\bibnamefont {Skordis}},\ }\href
  {\doibase 10.1016/j.physrep.2012.01.001} {\bibfield  {journal} {\bibinfo
  {journal} {Phys. Rept.}\ }\textbf {\bibinfo {volume} {513}},\ \bibinfo
  {pages} {1} (\bibinfo {year} {2012})},\ \Eprint
  {http://arxiv.org/abs/1106.2476} {arXiv:1106.2476 [astro-ph.CO]} \BibitemShut
  {NoStop}%
\bibitem [{\citenamefont {Tsujikawa}(2010)}]{Tsujikawa:2010zza}%
  \BibitemOpen
  \bibfield  {author} {\bibinfo {author} {\bibfnamefont {S.}~\bibnamefont
  {Tsujikawa}},\ }\href {\doibase 10.1007/978-3-642-10598-2_3} {\bibfield
  {journal} {\bibinfo  {journal} {Lect. Notes Phys.}\ }\textbf {\bibinfo
  {volume} {800}},\ \bibinfo {pages} {99} (\bibinfo {year} {2010})},\ \Eprint
  {http://arxiv.org/abs/1101.0191} {arXiv:1101.0191 [gr-qc]} \BibitemShut
  {NoStop}%
\bibitem [{\citenamefont {Rovelli}(2004)}]{Rovelli:2004tv}%
  \BibitemOpen
  \bibfield  {author} {\bibinfo {author} {\bibfnamefont {C.}~\bibnamefont
  {Rovelli}},\ }\href {\doibase 10.1017/CBO9780511755804} {\emph {\bibinfo
  {title} {{Quantum gravity}}}},\ Cambridge Monographs on Mathematical Physics\
  (\bibinfo  {publisher} {Univ. Pr.},\ \bibinfo {address} {Cambridge, UK},\
  \bibinfo {year} {2004})\BibitemShut {NoStop}%
\bibitem [{\citenamefont {Ashtekar}\ and\ \citenamefont
  {Lewandowski}(2004)}]{Ashtekar:2004eh}%
  \BibitemOpen
  \bibfield  {author} {\bibinfo {author} {\bibfnamefont {A.}~\bibnamefont
  {Ashtekar}}\ and\ \bibinfo {author} {\bibfnamefont {J.}~\bibnamefont
  {Lewandowski}},\ }\href {\doibase 10.1088/0264-9381/21/15/R01} {\bibfield
  {journal} {\bibinfo  {journal} {Class. Quant. Grav.}\ }\textbf {\bibinfo
  {volume} {21}},\ \bibinfo {pages} {R53} (\bibinfo {year} {2004})},\ \Eprint
  {http://arxiv.org/abs/gr-qc/0404018} {arXiv:gr-qc/0404018} \BibitemShut
  {NoStop}%
\bibitem [{\citenamefont {{Nicolis}}\ \emph {et~al.}(2009)\citenamefont
  {{Nicolis}}, \citenamefont {{Rattazzi}},\ and\ \citenamefont
  {{Trincherini}}}]{2009PhRvD..79f4036N}%
  \BibitemOpen
  \bibfield  {author} {\bibinfo {author} {\bibfnamefont {A.}~\bibnamefont
  {{Nicolis}}}, \bibinfo {author} {\bibfnamefont {R.}~\bibnamefont
  {{Rattazzi}}}, \ and\ \bibinfo {author} {\bibfnamefont {E.}~\bibnamefont
  {{Trincherini}}},\ }\href {\doibase 10.1103/PhysRevD.79.064036} {\bibfield
  {journal} {\bibinfo  {journal} {\prd}\ }\textbf {\bibinfo {volume} {79}},\
  \bibinfo {eid} {064036} (\bibinfo {year} {2009})},\ \Eprint
  {http://arxiv.org/abs/0811.2197} {arXiv:0811.2197 [hep-th]} \BibitemShut
  {NoStop}%
\bibitem [{\citenamefont {{Deffayet}}\ \emph
  {et~al.}(2009{\natexlab{a}})\citenamefont {{Deffayet}}, \citenamefont
  {{Esposito-Far{\`e}se}},\ and\ \citenamefont
  {{Vikman}}}]{2009PhRvD..79h4003D}%
  \BibitemOpen
  \bibfield  {author} {\bibinfo {author} {\bibfnamefont {C.}~\bibnamefont
  {{Deffayet}}}, \bibinfo {author} {\bibfnamefont {G.}~\bibnamefont
  {{Esposito-Far{\`e}se}}}, \ and\ \bibinfo {author} {\bibfnamefont
  {A.}~\bibnamefont {{Vikman}}},\ }\href {\doibase 10.1103/PhysRevD.79.084003}
  {\bibfield  {journal} {\bibinfo  {journal} {\prd}\ }\textbf {\bibinfo
  {volume} {79}},\ \bibinfo {eid} {084003} (\bibinfo {year}
  {2009}{\natexlab{a}})},\ \Eprint {http://arxiv.org/abs/0901.1314}
  {arXiv:0901.1314 [hep-th]} \BibitemShut {NoStop}%
\bibitem [{\citenamefont {{Deffayet}}\ \emph
  {et~al.}(2009{\natexlab{b}})\citenamefont {{Deffayet}}, \citenamefont
  {{Deser}},\ and\ \citenamefont
  {{Esposito-Far{\`e}se}}}]{2009PhRvD..80f4015D}%
  \BibitemOpen
  \bibfield  {author} {\bibinfo {author} {\bibfnamefont {C.}~\bibnamefont
  {{Deffayet}}}, \bibinfo {author} {\bibfnamefont {S.}~\bibnamefont {{Deser}}},
  \ and\ \bibinfo {author} {\bibfnamefont {G.}~\bibnamefont
  {{Esposito-Far{\`e}se}}},\ }\href {\doibase 10.1103/PhysRevD.80.064015}
  {\bibfield  {journal} {\bibinfo  {journal} {\prd}\ }\textbf {\bibinfo
  {volume} {80}},\ \bibinfo {eid} {064015} (\bibinfo {year}
  {2009}{\natexlab{b}})},\ \Eprint {http://arxiv.org/abs/0906.1967}
  {arXiv:0906.1967 [gr-qc]} \BibitemShut {NoStop}%
\bibitem [{\citenamefont {Deffayet}\ \emph {et~al.}(2011)\citenamefont
  {Deffayet}, \citenamefont {Gao}, \citenamefont {Steer},\ and\ \citenamefont
  {Zahariade}}]{Deffayet:2011gz}%
  \BibitemOpen
  \bibfield  {author} {\bibinfo {author} {\bibfnamefont {C.}~\bibnamefont
  {Deffayet}}, \bibinfo {author} {\bibfnamefont {X.}~\bibnamefont {Gao}},
  \bibinfo {author} {\bibfnamefont {D.~A.}\ \bibnamefont {Steer}}, \ and\
  \bibinfo {author} {\bibfnamefont {G.}~\bibnamefont {Zahariade}},\ }\href
  {\doibase 10.1103/PhysRevD.84.064039} {\bibfield  {journal} {\bibinfo
  {journal} {Phys. Rev.}\ }\textbf {\bibinfo {volume} {D84}},\ \bibinfo {pages}
  {064039} (\bibinfo {year} {2011})},\ \Eprint {http://arxiv.org/abs/1103.3260}
  {arXiv:1103.3260 [hep-th]} \BibitemShut {NoStop}%
\bibitem [{\citenamefont {Heisenberg}(2014)}]{Heisenberg:2014rta}%
  \BibitemOpen
  \bibfield  {author} {\bibinfo {author} {\bibfnamefont {L.}~\bibnamefont
  {Heisenberg}},\ }\href {\doibase 10.1088/1475-7516/2014/05/015} {\bibfield
  {journal} {\bibinfo  {journal} {JCAP}\ }\textbf {\bibinfo {volume} {1405}},\
  \bibinfo {pages} {015} (\bibinfo {year} {2014})},\ \Eprint
  {http://arxiv.org/abs/1402.7026} {arXiv:1402.7026 [hep-th]} \BibitemShut
  {NoStop}%
\bibitem [{\citenamefont {Tasinato}(2014)}]{Tasinato:2014eka}%
  \BibitemOpen
  \bibfield  {author} {\bibinfo {author} {\bibfnamefont {G.}~\bibnamefont
  {Tasinato}},\ }\href {\doibase 10.1007/JHEP04(2014)067} {\bibfield  {journal}
  {\bibinfo  {journal} {JHEP}\ }\textbf {\bibinfo {volume} {1404}},\ \bibinfo
  {pages} {067} (\bibinfo {year} {2014})},\ \Eprint
  {http://arxiv.org/abs/1402.6450} {arXiv:1402.6450 [hep-th]} \BibitemShut
  {NoStop}%
\bibitem [{\citenamefont {Allys}\ \emph
  {et~al.}(2016{\natexlab{a}})\citenamefont {Allys}, \citenamefont {Peter},\
  and\ \citenamefont {Rodr\'{\i}guez}}]{Allys:2015sht}%
  \BibitemOpen
  \bibfield  {author} {\bibinfo {author} {\bibfnamefont {E.}~\bibnamefont
  {Allys}}, \bibinfo {author} {\bibfnamefont {P.}~\bibnamefont {Peter}}, \ and\
  \bibinfo {author} {\bibfnamefont {Y.}~\bibnamefont {Rodr\'{\i}guez}},\ }\href
  {\doibase 10.1088/1475-7516/2016/02/004} {\bibfield  {journal} {\bibinfo
  {journal} {JCAP}\ }\textbf {\bibinfo {volume} {1602}},\ \bibinfo {pages}
  {004} (\bibinfo {year} {2016}{\natexlab{a}})},\ \Eprint
  {http://arxiv.org/abs/1511.03101} {arXiv:1511.03101 [hep-th]} \BibitemShut
  {NoStop}%
\bibitem [{\citenamefont {Beltr\'an~Jimenez}\ and\ \citenamefont
  {Heisenberg}(2016)}]{Jimenez:2016isa}%
  \BibitemOpen
  \bibfield  {author} {\bibinfo {author} {\bibfnamefont {J.}~\bibnamefont
  {Beltr\'an~Jimenez}}\ and\ \bibinfo {author} {\bibfnamefont {L.}~\bibnamefont
  {Heisenberg}},\ }\href {\doibase 10.1016/j.physletb.2016.04.017} {\bibfield
  {journal} {\bibinfo  {journal} {Phys. Lett.}\ }\textbf {\bibinfo {volume}
  {B757}},\ \bibinfo {pages} {405} (\bibinfo {year} {2016})},\ \Eprint
  {http://arxiv.org/abs/1602.03410} {arXiv:1602.03410 [hep-th]} \BibitemShut
  {NoStop}%
\bibitem [{\citenamefont {Allys}\ \emph
  {et~al.}(2016{\natexlab{b}})\citenamefont {Allys}, \citenamefont
  {Beltr\'an~Almeida}, \citenamefont {Peter},\ and\ \citenamefont
  {Rodr\'{\i}guez}}]{Allys:2016jaq}%
  \BibitemOpen
  \bibfield  {author} {\bibinfo {author} {\bibfnamefont {E.}~\bibnamefont
  {Allys}}, \bibinfo {author} {\bibfnamefont {J.~P.}\ \bibnamefont
  {Beltr\'an~Almeida}}, \bibinfo {author} {\bibfnamefont {P.}~\bibnamefont
  {Peter}}, \ and\ \bibinfo {author} {\bibfnamefont {Y.}~\bibnamefont
  {Rodr\'{\i}guez}},\ }\href {\doibase 10.1088/1475-7516/2016/09/026}
  {\bibfield  {journal} {\bibinfo  {journal} {JCAP}\ }\textbf {\bibinfo
  {volume} {1609}},\ \bibinfo {pages} {026} (\bibinfo {year}
  {2016}{\natexlab{b}})},\ \Eprint {http://arxiv.org/abs/1605.08355}
  {arXiv:1605.08355 [hep-th]} \BibitemShut {NoStop}%
\bibitem [{\citenamefont {Kobayashi}\ \emph {et~al.}(2011)\citenamefont
  {Kobayashi}, \citenamefont {Yamaguchi},\ and\ \citenamefont
  {Yokoyama}}]{Kobayashi:2011nu}%
  \BibitemOpen
  \bibfield  {author} {\bibinfo {author} {\bibfnamefont {T.}~\bibnamefont
  {Kobayashi}}, \bibinfo {author} {\bibfnamefont {M.}~\bibnamefont
  {Yamaguchi}}, \ and\ \bibinfo {author} {\bibfnamefont {J.}~\bibnamefont
  {Yokoyama}},\ }\href {\doibase 10.1143/PTP.126.511} {\bibfield  {journal}
  {\bibinfo  {journal} {Prog. Theor. Phys.}\ }\textbf {\bibinfo {volume}
  {126}},\ \bibinfo {pages} {511} (\bibinfo {year} {2011})},\ \Eprint
  {http://arxiv.org/abs/1105.5723} {arXiv:1105.5723 [hep-th]} \BibitemShut
  {NoStop}%
\bibitem [{\citenamefont {Horndeski}(1974)}]{Horndeski:1974wa}%
  \BibitemOpen
  \bibfield  {author} {\bibinfo {author} {\bibfnamefont {G.~W.}\ \bibnamefont
  {Horndeski}},\ }\href {\doibase 10.1007/BF01807638} {\bibfield  {journal}
  {\bibinfo  {journal} {Int. J. Theor. Phys.}\ }\textbf {\bibinfo {volume}
  {10}},\ \bibinfo {pages} {363} (\bibinfo {year} {1974})}\BibitemShut
  {NoStop}%
\bibitem [{\citenamefont {Horndeski}(1976)}]{Horndeski:1976gi}%
  \BibitemOpen
  \bibfield  {author} {\bibinfo {author} {\bibfnamefont {G.~W.}\ \bibnamefont
  {Horndeski}},\ }\href {\doibase 10.1063/1.522837} {\bibfield  {journal}
  {\bibinfo  {journal} {J. Math. Phys.}\ }\textbf {\bibinfo {volume} {17}},\
  \bibinfo {pages} {1980} (\bibinfo {year} {1976})}\BibitemShut {NoStop}%
\bibitem [{\citenamefont {Ostrogradsky}(1850)}]{Ostrogradsky:1850fid}%
  \BibitemOpen
  \bibfield  {author} {\bibinfo {author} {\bibfnamefont {M.}~\bibnamefont
  {Ostrogradsky}},\ }\href@noop {} {\bibfield  {journal} {\bibinfo  {journal}
  {Mem. Acad. St. Petersbourg}\ }\textbf {\bibinfo {volume} {6}},\ \bibinfo
  {pages} {385} (\bibinfo {year} {1850})}\BibitemShut {NoStop}%
\bibitem [{\citenamefont {Woodard}(2007)}]{Woodard:2006nt}%
  \BibitemOpen
  \bibfield  {author} {\bibinfo {author} {\bibfnamefont {R.~P.}\ \bibnamefont
  {Woodard}},\ }\bibfield  {booktitle} {\emph {\bibinfo {booktitle} {{The
  invisible universe: Dark matter and dark energy. Proceedings, 3rd Aegean
  School, Karfas, Greece, September 26-October 1, 2005}}},\ }\href {\doibase
  10.1007/978-3-540-71013-4_14} {\bibfield  {journal} {\bibinfo  {journal}
  {Lect. Notes Phys.}\ }\textbf {\bibinfo {volume} {720}},\ \bibinfo {pages}
  {403} (\bibinfo {year} {2007})},\ \Eprint
  {http://arxiv.org/abs/astro-ph/0601672} {arXiv:astro-ph/0601672} \BibitemShut
  {NoStop}%
\bibitem [{\citenamefont {Woodard}(2015)}]{Woodard:2015zca}%
  \BibitemOpen
  \bibfield  {author} {\bibinfo {author} {\bibfnamefont {R.~P.}\ \bibnamefont
  {Woodard}},\ }\href {\doibase 10.4249/scholarpedia.32243} {\bibfield
  {journal} {\bibinfo  {journal} {Scholarpedia}\ }\textbf {\bibinfo {volume}
  {10}},\ \bibinfo {pages} {32243} (\bibinfo {year} {2015})},\ \Eprint
  {http://arxiv.org/abs/1506.02210} {arXiv:1506.02210 [hep-th]} \BibitemShut
  {NoStop}%
\bibitem [{\citenamefont {Kobayashi}(2019)}]{Kobayashi:2019hrl}%
  \BibitemOpen
  \bibfield  {author} {\bibinfo {author} {\bibfnamefont {T.}~\bibnamefont
  {Kobayashi}},\ }\href {\doibase 10.1088/1361-6633/ab2429} {\bibfield
  {journal} {\bibinfo  {journal} {Rept. Prog. Phys.}\ }\textbf {\bibinfo
  {volume} {82}},\ \bibinfo {pages} {086901} (\bibinfo {year} {2019})},\
  \Eprint {http://arxiv.org/abs/1901.07183} {arXiv:1901.07183 [gr-qc]}
  \BibitemShut {NoStop}%
\bibitem [{\citenamefont {{Heisenberg}}(2019)}]{2019PhR...796....1H}%
  \BibitemOpen
  \bibfield  {author} {\bibinfo {author} {\bibfnamefont {L.}~\bibnamefont
  {{Heisenberg}}},\ }\href {\doibase 10.1016/j.physrep.2018.11.006} {\bibfield
  {journal} {\bibinfo  {journal} {\physrep}\ }\textbf {\bibinfo {volume}
  {796}},\ \bibinfo {pages} {1} (\bibinfo {year} {2019})},\ \Eprint
  {http://arxiv.org/abs/1807.01725} {arXiv:1807.01725 [gr-qc]} \BibitemShut
  {NoStop}%
\bibitem [{\citenamefont {{Allys}}\ \emph {et~al.}(2016)\citenamefont
  {{Allys}}, \citenamefont {{Peter}},\ and\ \citenamefont
  {{Rodr{\'\i}guez}}}]{2016PhRvD..94h4041A}%
  \BibitemOpen
  \bibfield  {author} {\bibinfo {author} {\bibfnamefont {E.}~\bibnamefont
  {{Allys}}}, \bibinfo {author} {\bibfnamefont {P.}~\bibnamefont {{Peter}}}, \
  and\ \bibinfo {author} {\bibfnamefont {Y.}~\bibnamefont {{Rodr{\'\i}guez}}},\
  }\href {\doibase 10.1103/PhysRevD.94.084041} {\bibfield  {journal} {\bibinfo
  {journal} {\prd}\ }\textbf {\bibinfo {volume} {94}},\ \bibinfo {eid} {084041}
  (\bibinfo {year} {2016})},\ \Eprint {http://arxiv.org/abs/1609.05870}
  {arXiv:1609.05870 [hep-th]} \BibitemShut {NoStop}%
\bibitem [{\citenamefont {Beltr\'an~Jimenez}\ and\ \citenamefont
  {Heisenberg}(2017)}]{Jimenez:2016upj}%
  \BibitemOpen
  \bibfield  {author} {\bibinfo {author} {\bibfnamefont {J.}~\bibnamefont
  {Beltr\'an~Jimenez}}\ and\ \bibinfo {author} {\bibfnamefont {L.}~\bibnamefont
  {Heisenberg}},\ }\href {\doibase 10.1016/j.physletb.2017.03.002} {\bibfield
  {journal} {\bibinfo  {journal} {Phys. Lett.}\ }\textbf {\bibinfo {volume}
  {B770}},\ \bibinfo {pages} {16} (\bibinfo {year} {2017})},\ \Eprint
  {http://arxiv.org/abs/1610.08960} {arXiv:1610.08960 [hep-th]} \BibitemShut
  {NoStop}%
\bibitem [{\citenamefont
  {Armend\'ariz-Pic\'on}(2004)}]{ArmendarizPicon:2004pm}%
  \BibitemOpen
  \bibfield  {author} {\bibinfo {author} {\bibfnamefont {C.}~\bibnamefont
  {Armend\'ariz-Pic\'on}},\ }\href {\doibase 10.1088/1475-7516/2004/07/007}
  {\bibfield  {journal} {\bibinfo  {journal} {JCAP}\ }\textbf {\bibinfo
  {volume} {0407}},\ \bibinfo {pages} {007} (\bibinfo {year} {2004})},\ \Eprint
  {http://arxiv.org/abs/astro-ph/0405267} {arXiv:astro-ph/0405267} \BibitemShut
  {NoStop}%
\bibitem [{\citenamefont {Maleknejad}\ and\ \citenamefont
  {Sheikh-Jabbari}(2013)}]{Maleknejad:2011jw}%
  \BibitemOpen
  \bibfield  {author} {\bibinfo {author} {\bibfnamefont {A.}~\bibnamefont
  {Maleknejad}}\ and\ \bibinfo {author} {\bibfnamefont {M.~M.}\ \bibnamefont
  {Sheikh-Jabbari}},\ }\href {\doibase 10.1016/j.physletb.2013.05.001}
  {\bibfield  {journal} {\bibinfo  {journal} {Phys. Lett.}\ }\textbf {\bibinfo
  {volume} {B723}},\ \bibinfo {pages} {224} (\bibinfo {year} {2013})},\ \Eprint
  {http://arxiv.org/abs/1102.1513} {arXiv:1102.1513 [hep-ph]} \BibitemShut
  {NoStop}%
\bibitem [{\citenamefont {{Maleknejad}}\ and\ \citenamefont
  {{Sheikh-Jabbari}}(2011)}]{2011PhRvD..84d3515M}%
  \BibitemOpen
  \bibfield  {author} {\bibinfo {author} {\bibfnamefont {A.}~\bibnamefont
  {{Maleknejad}}}\ and\ \bibinfo {author} {\bibfnamefont {M.~M.}\ \bibnamefont
  {{Sheikh-Jabbari}}},\ }\href {\doibase 10.1103/PhysRevD.84.043515} {\bibfield
   {journal} {\bibinfo  {journal} {\prd}\ }\textbf {\bibinfo {volume} {84}},\
  \bibinfo {eid} {043515} (\bibinfo {year} {2011})},\ \Eprint
  {http://arxiv.org/abs/1102.1932} {arXiv:1102.1932 [hep-ph]} \BibitemShut
  {NoStop}%
\bibitem [{\citenamefont {Nieto}\ and\ \citenamefont
  {Rodr\'{\i}guez}(2016)}]{Nieto:2016gnp}%
  \BibitemOpen
  \bibfield  {author} {\bibinfo {author} {\bibfnamefont {C.~M.}\ \bibnamefont
  {Nieto}}\ and\ \bibinfo {author} {\bibfnamefont {Y.}~\bibnamefont
  {Rodr\'{\i}guez}},\ }\href {\doibase 10.1142/S0217732316400058} {\bibfield
  {journal} {\bibinfo  {journal} {Mod. Phys. Lett.}\ }\textbf {\bibinfo
  {volume} {A31}},\ \bibinfo {pages} {1640005} (\bibinfo {year} {2016})},\
  \Eprint {http://arxiv.org/abs/1602.07197} {arXiv:1602.07197 [gr-qc]}
  \BibitemShut {NoStop}%
\bibitem [{\citenamefont {Adshead}\ and\ \citenamefont
  {Sfakianakis}(2017)}]{Adshead:2017hnc}%
  \BibitemOpen
  \bibfield  {author} {\bibinfo {author} {\bibfnamefont {P.}~\bibnamefont
  {Adshead}}\ and\ \bibinfo {author} {\bibfnamefont {E.~I.}\ \bibnamefont
  {Sfakianakis}},\ }\href {\doibase 10.1007/JHEP08(2017)130} {\bibfield
  {journal} {\bibinfo  {journal} {JHEP}\ }\textbf {\bibinfo {volume} {1708}},\
  \bibinfo {pages} {130} (\bibinfo {year} {2017})},\ \Eprint
  {http://arxiv.org/abs/1705.03024} {arXiv:1705.03024 [hep-th]} \BibitemShut
  {NoStop}%
\bibitem [{\citenamefont {{Adshead}}\ and\ \citenamefont
  {{Wyman}}(2012)}]{2012PhRvL.108z1302A}%
  \BibitemOpen
  \bibfield  {author} {\bibinfo {author} {\bibfnamefont {P.}~\bibnamefont
  {{Adshead}}}\ and\ \bibinfo {author} {\bibfnamefont {M.}~\bibnamefont
  {{Wyman}}},\ }\href {\doibase 10.1103/PhysRevLett.108.261302} {\bibfield
  {journal} {\bibinfo  {journal} {\prl}\ }\textbf {\bibinfo {volume} {108}},\
  \bibinfo {eid} {261302} (\bibinfo {year} {2012})},\ \Eprint
  {http://arxiv.org/abs/1202.2366} {arXiv:1202.2366 [hep-th]} \BibitemShut
  {NoStop}%
\bibitem [{\citenamefont {Adshead}\ \emph {et~al.}(2016)\citenamefont
  {Adshead}, \citenamefont {Martinec}, \citenamefont {Sfakianakis},\ and\
  \citenamefont {Wyman}}]{Adshead:2016omu}%
  \BibitemOpen
  \bibfield  {author} {\bibinfo {author} {\bibfnamefont {P.}~\bibnamefont
  {Adshead}}, \bibinfo {author} {\bibfnamefont {E.}~\bibnamefont {Martinec}},
  \bibinfo {author} {\bibfnamefont {E.~I.}\ \bibnamefont {Sfakianakis}}, \ and\
  \bibinfo {author} {\bibfnamefont {M.}~\bibnamefont {Wyman}},\ }\href
  {\doibase 10.1007/JHEP12(2016)137} {\bibfield  {journal} {\bibinfo  {journal}
  {JHEP}\ }\textbf {\bibinfo {volume} {1612}},\ \bibinfo {pages} {137}
  (\bibinfo {year} {2016})},\ \Eprint {http://arxiv.org/abs/1609.04025}
  {arXiv:1609.04025 [hep-th]} \BibitemShut {NoStop}%
\bibitem [{\citenamefont {\'Alvarez}\ \emph {et~al.}(2019)\citenamefont
  {\'Alvarez}, \citenamefont {Orjuela-Quintana}, \citenamefont
  {Rodr\'{\i}guez},\ and\ \citenamefont {Valenzuela-Toledo}}]{Alvarez:2019ues}%
  \BibitemOpen
  \bibfield  {author} {\bibinfo {author} {\bibfnamefont {M.}~\bibnamefont
  {\'Alvarez}}, \bibinfo {author} {\bibfnamefont {J.~B.}\ \bibnamefont
  {Orjuela-Quintana}}, \bibinfo {author} {\bibfnamefont {Y.}~\bibnamefont
  {Rodr\'{\i}guez}}, \ and\ \bibinfo {author} {\bibfnamefont {C.~A.}\
  \bibnamefont {Valenzuela-Toledo}},\ }\href {\doibase
  10.1088/1361-6382/ab3775} {\bibfield  {journal} {\bibinfo  {journal} {Class.
  Quant. Grav.}\ }\textbf {\bibinfo {volume} {36}},\ \bibinfo {pages} {195004}
  (\bibinfo {year} {2019})},\ \Eprint {http://arxiv.org/abs/1901.04624}
  {arXiv:1901.04624 [gr-qc]} \BibitemShut {NoStop}%
\bibitem [{\citenamefont {Emami}\ \emph {et~al.}(2017)\citenamefont {Emami},
  \citenamefont {Mukohyama}, \citenamefont {Namba},\ and\ \citenamefont
  {Zhang}}]{Emami:2016ldl}%
  \BibitemOpen
  \bibfield  {author} {\bibinfo {author} {\bibfnamefont {R.}~\bibnamefont
  {Emami}}, \bibinfo {author} {\bibfnamefont {S.}~\bibnamefont {Mukohyama}},
  \bibinfo {author} {\bibfnamefont {R.}~\bibnamefont {Namba}}, \ and\ \bibinfo
  {author} {\bibfnamefont {Y.-l.}\ \bibnamefont {Zhang}},\ }\href {\doibase
  10.1088/1475-7516/2017/03/058} {\bibfield  {journal} {\bibinfo  {journal}
  {JCAP}\ }\textbf {\bibinfo {volume} {1703}},\ \bibinfo {pages} {058}
  (\bibinfo {year} {2017})},\ \Eprint {http://arxiv.org/abs/1612.09581}
  {arXiv:1612.09581 [hep-th]} \BibitemShut {NoStop}%
\bibitem [{\citenamefont {Rodr\'{\i}guez}\ and\ \citenamefont
  {Navarro}(2017)}]{Rodriguez:2017ckc}%
  \BibitemOpen
  \bibfield  {author} {\bibinfo {author} {\bibfnamefont {Y.}~\bibnamefont
  {Rodr\'{\i}guez}}\ and\ \bibinfo {author} {\bibfnamefont {A.~A.}\
  \bibnamefont {Navarro}},\ }\bibfield  {booktitle} {\emph {\bibinfo
  {booktitle} {{Proceedings, 70\&70 Classical and Quantum Gravitation Party:
  Meeting with Two Latin American Masters on Theoretical Physics: Cartagena,
  Colombia, September 28-30, 2016}}},\ }\href {\doibase
  10.1088/1742-6596/831/1/012004} {\bibfield  {journal} {\bibinfo  {journal}
  {J. Phys. Conf. Ser.}\ }\textbf {\bibinfo {volume} {831}},\ \bibinfo {pages}
  {012004} (\bibinfo {year} {2017})},\ \Eprint
  {http://arxiv.org/abs/1703.01884} {arXiv:1703.01884 [hep-th]} \BibitemShut
  {NoStop}%
\bibitem [{\citenamefont {Rodr\'{\i}guez}\ and\ \citenamefont
  {Navarro}(2018)}]{Rodriguez:2017wkg}%
  \BibitemOpen
  \bibfield  {author} {\bibinfo {author} {\bibfnamefont {Y.}~\bibnamefont
  {Rodr\'{\i}guez}}\ and\ \bibinfo {author} {\bibfnamefont {A.~A.}\
  \bibnamefont {Navarro}},\ }\href {\doibase 10.1016/j.dark.2018.01.003}
  {\bibfield  {journal} {\bibinfo  {journal} {Phys. Dark Univ.}\ }\textbf
  {\bibinfo {volume} {19}},\ \bibinfo {pages} {129} (\bibinfo {year} {2018})},\
  \Eprint {http://arxiv.org/abs/1711.01935} {arXiv:1711.01935 [gr-qc]}
  \BibitemShut {NoStop}%
\bibitem [{\citenamefont {Garnica}\ \emph {et~al.}(2019)\citenamefont
  {Garnica}, \citenamefont {G\'omez}, \citenamefont {Navarro},\ and\
  \citenamefont {Rodr\'{\i}guez}}]{infprep}%
  \BibitemOpen
  \bibfield  {author} {\bibinfo {author} {\bibfnamefont {J.~C.}\ \bibnamefont
  {Garnica}}, \bibinfo {author} {\bibfnamefont {L.~G.}\ \bibnamefont
  {G\'omez}}, \bibinfo {author} {\bibfnamefont {A.~A.}\ \bibnamefont
  {Navarro}}, \ and\ \bibinfo {author} {\bibfnamefont {Y.}~\bibnamefont
  {Rodr\'{\i}guez}},\ }\href@noop {} {\bibfield  {journal} {\bibinfo  {journal}
  {Work in progress}\ } (\bibinfo {year} {2019})}\BibitemShut {NoStop}%
\bibitem [{\citenamefont {Gallego~Cadavid}\ and\ \citenamefont
  {Rodr\'{\i}guez}(2019)}]{beyondsu2}%
  \BibitemOpen
  \bibfield  {author} {\bibinfo {author} {\bibfnamefont {A.}~\bibnamefont
  {Gallego~Cadavid}}\ and\ \bibinfo {author} {\bibfnamefont {Y.}~\bibnamefont
  {Rodr\'{\i}guez}},\ }\href@noop {} {\bibfield  {journal} {\bibinfo  {journal}
  {Work in progress}\ } (\bibinfo {year} {2019})}\BibitemShut {NoStop}%
\bibitem [{\citenamefont {Errasti~D\'{\i}ez}\ \emph
  {et~al.}(2019{\natexlab{a}})\citenamefont {Errasti~D\'{\i}ez}, \citenamefont
  {Gording}, \citenamefont {M\'endez-Zavaleta},\ and\ \citenamefont
  {Schmidt-May}}]{ErrastiDiez:2019ttn}%
  \BibitemOpen
  \bibfield  {author} {\bibinfo {author} {\bibfnamefont {V.}~\bibnamefont
  {Errasti~D\'{\i}ez}}, \bibinfo {author} {\bibfnamefont {B.}~\bibnamefont
  {Gording}}, \bibinfo {author} {\bibfnamefont {J.~A.}\ \bibnamefont
  {M\'endez-Zavaleta}}, \ and\ \bibinfo {author} {\bibfnamefont
  {A.}~\bibnamefont {Schmidt-May}},\ }\href@noop {} {\  (\bibinfo {year}
  {2019}{\natexlab{a}})},\ \Eprint {http://arxiv.org/abs/1905.06967}
  {arXiv:1905.06967 [hep-th]} \BibitemShut {NoStop}%
\bibitem [{\citenamefont {Errasti~D\'{\i}ez}\ \emph
  {et~al.}(2019{\natexlab{b}})\citenamefont {Errasti~D\'{\i}ez}, \citenamefont
  {Gording}, \citenamefont {M\'endez-Zavaleta},\ and\ \citenamefont
  {Schmidt-May}}]{ErrastiDiez:2019trb}%
  \BibitemOpen
  \bibfield  {author} {\bibinfo {author} {\bibfnamefont {V.}~\bibnamefont
  {Errasti~D\'{\i}ez}}, \bibinfo {author} {\bibfnamefont {B.}~\bibnamefont
  {Gording}}, \bibinfo {author} {\bibfnamefont {J.~A.}\ \bibnamefont
  {M\'endez-Zavaleta}}, \ and\ \bibinfo {author} {\bibfnamefont
  {A.}~\bibnamefont {Schmidt-May}},\ }\href@noop {} {\  (\bibinfo {year}
  {2019}{\natexlab{b}})},\ \Eprint {http://arxiv.org/abs/1905.06968}
  {arXiv:1905.06968 [hep-th]} \BibitemShut {NoStop}%
\bibitem [{\citenamefont {{Caldwell}}\ \emph {et~al.}(2016)\citenamefont
  {{Caldwell}}, \citenamefont {{Devulder}},\ and\ \citenamefont
  {{Maksimova}}}]{2016PhRvD..94f3005C}%
  \BibitemOpen
  \bibfield  {author} {\bibinfo {author} {\bibfnamefont {R.~R.}\ \bibnamefont
  {{Caldwell}}}, \bibinfo {author} {\bibfnamefont {C.}~\bibnamefont
  {{Devulder}}}, \ and\ \bibinfo {author} {\bibfnamefont {N.~A.}\ \bibnamefont
  {{Maksimova}}},\ }\href {\doibase 10.1103/PhysRevD.94.063005} {\bibfield
  {journal} {\bibinfo  {journal} {\prd}\ }\textbf {\bibinfo {volume} {94}},\
  \bibinfo {eid} {063005} (\bibinfo {year} {2016})},\ \Eprint
  {http://arxiv.org/abs/1604.08939} {arXiv:1604.08939 [gr-qc]} \BibitemShut
  {NoStop}%
\bibitem [{\citenamefont {Caldwell}\ and\ \citenamefont
  {Devulder}(2018)}]{Caldwell:2018feo}%
  \BibitemOpen
  \bibfield  {author} {\bibinfo {author} {\bibfnamefont {R.~R.}\ \bibnamefont
  {Caldwell}}\ and\ \bibinfo {author} {\bibfnamefont {C.}~\bibnamefont
  {Devulder}},\ }\href@noop {} {\  (\bibinfo {year} {2018})},\ \Eprint
  {http://arxiv.org/abs/1802.07371} {arXiv:1802.07371 [gr-qc]} \BibitemShut
  {NoStop}%
\bibitem [{\citenamefont {Sbis\`a}(2015)}]{Sbisa:2014pzo}%
  \BibitemOpen
  \bibfield  {author} {\bibinfo {author} {\bibfnamefont {F.}~\bibnamefont
  {Sbis\`a}},\ }\href {\doibase 10.1088/0143-0807/36/1/015009} {\bibfield
  {journal} {\bibinfo  {journal} {Eur. J. Phys.}\ }\textbf {\bibinfo {volume}
  {36}},\ \bibinfo {pages} {015009} (\bibinfo {year} {2015})},\ \Eprint
  {http://arxiv.org/abs/1406.4550} {arXiv:1406.4550 [hep-th]} \BibitemShut
  {NoStop}%
\bibitem [{\citenamefont {Babichev}\ \emph {et~al.}(2008)\citenamefont
  {Babichev}, \citenamefont {Mukhanov},\ and\ \citenamefont
  {Vikman}}]{Babichev:2007dw}%
  \BibitemOpen
  \bibfield  {author} {\bibinfo {author} {\bibfnamefont {E.}~\bibnamefont
  {Babichev}}, \bibinfo {author} {\bibfnamefont {V.}~\bibnamefont {Mukhanov}},
  \ and\ \bibinfo {author} {\bibfnamefont {A.}~\bibnamefont {Vikman}},\ }\href
  {\doibase 10.1088/1126-6708/2008/02/101} {\bibfield  {journal} {\bibinfo
  {journal} {JHEP}\ }\textbf {\bibinfo {volume} {0802}},\ \bibinfo {pages}
  {101} (\bibinfo {year} {2008})},\ \Eprint {http://arxiv.org/abs/0708.0561}
  {arXiv:0708.0561 [hep-th]} \BibitemShut {NoStop}%
\bibitem [{\citenamefont {{Goon}}\ \emph {et~al.}(2011)\citenamefont {{Goon}},
  \citenamefont {{Hinterbichler}},\ and\ \citenamefont
  {{Trodden}}}]{2011PhRvD..83h5015G}%
  \BibitemOpen
  \bibfield  {author} {\bibinfo {author} {\bibfnamefont {G.~L.}\ \bibnamefont
  {{Goon}}}, \bibinfo {author} {\bibfnamefont {K.}~\bibnamefont
  {{Hinterbichler}}}, \ and\ \bibinfo {author} {\bibfnamefont {M.}~\bibnamefont
  {{Trodden}}},\ }\href {\doibase 10.1103/PhysRevD.83.085015} {\bibfield
  {journal} {\bibinfo  {journal} {\prd}\ }\textbf {\bibinfo {volume} {83}},\
  \bibinfo {eid} {085015} (\bibinfo {year} {2011})},\ \Eprint
  {http://arxiv.org/abs/1008.4580} {arXiv:1008.4580 [hep-th]} \BibitemShut
  {NoStop}%
\bibitem [{\citenamefont {de~Fromont}\ \emph {et~al.}(2013)\citenamefont
  {de~Fromont}, \citenamefont {de~Rham}, \citenamefont {Heisenberg},\ and\
  \citenamefont {Matas}}]{deFromont:2013iwa}%
  \BibitemOpen
  \bibfield  {author} {\bibinfo {author} {\bibfnamefont {P.}~\bibnamefont
  {de~Fromont}}, \bibinfo {author} {\bibfnamefont {C.}~\bibnamefont {de~Rham}},
  \bibinfo {author} {\bibfnamefont {L.}~\bibnamefont {Heisenberg}}, \ and\
  \bibinfo {author} {\bibfnamefont {A.}~\bibnamefont {Matas}},\ }\href
  {\doibase 10.1007/JHEP07(2013)067} {\bibfield  {journal} {\bibinfo  {journal}
  {JHEP}\ }\textbf {\bibinfo {volume} {1307}},\ \bibinfo {pages} {067}
  (\bibinfo {year} {2013})},\ \Eprint {http://arxiv.org/abs/1303.0274}
  {arXiv:1303.0274 [hep-th]} \BibitemShut {NoStop}%
\end{thebibliography}%
\bibliographystyle{apsrev4-1}
\end{document}